# A Fourier Neural Operator for Accelerated Discretization-Invariant Solutions of the Radiative Transfer Equation

Daniel Carne[a]

[a] Department of Mechanical and Nuclear Engineering, United States Naval Academy, Annapolis, MD 21402, USA

[*] Corresponding Author: carne@usna.edu

**Abstract**

The radiative transfer equation (RTE) governs thermal radiation in participating media, which is critical to modeling combustion, atmospheric, high-temperature, and radiative thermal management applications. However, due to the inherently high-dimensional nature of radiative transfer, numerical solutions to the RTE induce significant computational cost. This work develops a Fourier neural operator (FNO) as a fast, discretization-invariant, surrogate model for predicting the absorbed heat flux field in participating media. A dataset is generated on a benchmark 2-dimensional problem modeling thermal surface emission into a participating medium with spatially varying properties using Monte Carlo simulations. The FNO is trained to learn the solution operator, mapping the input property fields and geometry to the resulting heat flux field. The trained FNO provides up to 26× computational acceleration compared to Monte Carlo simulation at equivalent error. Furthermore, as the FNO learns the solution operator and not an image-to-image mapping, the trained FNO accurately generalizes to various spatial discretizations, including those not seen in the training dataset. A spectral analysis of the spatial frequencies present in the heat flux prediction demonstrates the FNO's ability to suppress the high-frequency noise present in the training dataset, while preserving the physically meaningful low-mid frequency spectrum. These results demonstrate the Fourier neural operator surrogate model can provide accurate, computationally efficient, and discretization-invariant predictions for radiative transfer in a participating medium. By learning the underlying solution operator, this approach moves toward the development of surrogate models for radiative transfer capable of generalizing across a broad range of geometries, discretizations, and boundary conditions.

## Introduction

Thermal radiative transport, involving scattering, absorption, and emission processes, plays a central role in many important applications and participating media such as combustion gases[1,2], atmospheric aerosols and clouds[3], high-temperature systems[4], and radiative thermal management[5,6]. In these systems, radiation is governed by the radiative transfer equation (RTE), which is a partial differential equation that quantifies the transport of radiant intensity[7]. Although the RTE provides a rigorous framework for modeling radiative transport in participating media, solving it accurately remains computationally expensive. Unlike conventional heat transfer and fluid dynamics equations, radiative transport is inherently high-dimensional. Solutions depend not only on spatial position, but also on propagation direction and wavelength. Deterministic methods such as the discrete ordinates method therefore require spatial, angular, and spectral discretization, and Monte Carlo stochastic methods require a significant number of photons modeled to reduce statistical noise across all dimensions[8]. These computational costs become especially limiting in coupled thermal-fluid problems, transient problems, design optimization, and inverse problems, where the RTE must be repeatedly solved. As a result, radiative transport is often simplified, neglected, or treated with a reduced-order approximation. Thus, faster solvers are necessary to preserve the fidelity of high-resolution RTE solutions while reducing computational burden.

To alleviate the computational cost of directly solving the RTE, machine learning has increasingly been explored as a means of accelerating radiative transport simulations. These approaches typically use simulation or experimental data to train surrogate models that predict quantities such as absorption fields, reflectance and transmittance, or effective optical properties. Once trained, these surrogate models enable accelerated predictions at a fraction of the cost of conventional RTE solvers, making them particularly attractive for applications that require repeated evaluations. Prior studies have explored a variety of machine learning architectures across a broad range of radiative transfer applications. For example, Kang et al. developed a fully connected neural network to predict transmission through metallic packed beds using three input features that describe the geometric parameters and material properties[9]. This type of parameter-based surrogate model has shown, across a variety of applications, to provide substantial acceleration because the network input is low-dimensional due to the fixed underlying geometry[10,11]. However, while this approach is highly efficient for specific applications, the trained models are not directly generalizable to different geometries or alternative problem formulations. Other approaches, such as by Peng et al., have used variants of convolutional neural networks and U-nets to denoise low-fidelity Monte Carlo predictions[12–15]. These methods are attractive because they retain the flexibility of the Monte Carlo method while reducing the number of photon histories required. However, because each prediction still requires an underlying Monte Carlo simulation, the computational cost remains tied to photon transport and can remain substantially higher than purely data-driven surrogate models that bypass the numerical RTE solve after training. Additionally, although convolutional architectures are not necessarily restricted to a single input size, they are commonly trained as image-to-image models whose performance depends on the spatial discretization and resolution used during training.

Neural operators provide a different approach for developing machine-learning surrogate models for solving partial differential equations[16,17]. Rather than learning mappings between fixed-dimensional arrays, neural operators learn mappings between functions. This is well suited for developing a generalizable model for the RTE, where both the inputs, such as optical properties, temperature, and source terms, and the outputs, such as absorption, intensity, and heat flux, are spatially varying fields. Thus, neural operators can approximate the RTE solution operator rather than an image-to-image correction or a fixed-geometry surrogate. This mesh-free formulation allows a single trained model to be evaluated on discretized domains with different spatial resolutions. Among neural operator architectures, Li et al. developed the Fourier neural operator (FNO), which parameterizes the integral operator using convolutional operations performed in Fourier space[18]. This enables FNOs to efficiently model nonlocal interactions and has been shown to provide high accuracy and significant acceleration compared with traditional PDE solvers. Variants of FNO architectures have since been applied to accelerate a broad range of physics-based problems, including fluid dynamics[19–21], heat transfer[22,23], solid mechanics[24,25], and other systems governed by partial differential equations. Applications specific to radiative transfer include Farmer et al. who applied an FNO for predicting the behavior of Marshak waves[26], Yao and Zhong et al. who applied an FNO for predicting atmospheric fluxes and heating rates[27], and Rost et al. who applied a U-FNO for predicting radiative intensity in astrophysics applications[28].

Existing machine learning approaches for radiative transfer are typically application specific. In contrast, the long-term aim of this work is the development of a general surrogate solver for radiative transfer that is independent of application. In this study, we develop and evaluate a Fourier neural operator architecture for accelerated, discretization-invariant prediction of radiative heat flux fields in participating media governed by the radiative transfer equation. The Monte Carlo method is used to generate training and testing data on a two-dimensional radiative transport benchmark problem with spatially varying absorption coefficient, scattering coefficient, asymmetry parameter, temperature, and geometry fields. The trained FNO learns a mapping from these input fields to the resulting heat flux field, thereby approximating the RTE solution operator rather than a fixed-resolution image-to-image surrogate. The model is evaluated by examining training-data scaling, network-width scaling, sensitivity to Monte Carlo photon count, prediction accuracy relative to direct Monte Carlo simulation, and computational acceleration at comparable error. In addition, the discretization-invariant character of the FNO is tested by applying models trained at one spatial resolution to finer and coarser grids without retraining. Finally, Fourier-domain spectral analyses are used to assess how the FNO represents physically relevant spatial structure while filtering high-frequency Monte Carlo noise. These results demonstrate the potential of Fourier neural operators as fast surrogate solvers for radiative transfer problems.

## Methodology

*Fourier neural operator & architecture*

The FNO framework implemented here is based on the study by Li et al.[18]. Let $D \subset \mathbb{R}^d$ be a bounded and open set, and let $A = A(D;\ \mathbb{R}^{d_a})$ and $U = U(D;\ \mathbb{R}^{d_u})$ define the input and output function spaces, respectively. Additionally, let $G^\dagger: A \to U$ be a non-linear map which maps the input fields to the corresponding solution of the radiative transfer equation (RTE). A dataset is generated where independent samples $\{a_j, u_j\}_{j=1}^N$ are collected where the observed outputs of $u_j = G^\dagger(a_j) + \varepsilon_j$ are corrupted with noise ($\varepsilon_j$) due to the Monte Carlo method used to approximate solutions to the RTE. We aim to develop an FNO which approximates $G^\dagger$ with $G_\theta: A \to U$, where $\theta \in \Theta$ is the network parameters, by minimizing the cost function ($C$):

$$\min_\theta \frac{1}{N} \sum_{j=1}^{N} C\left(G_\theta(a_j), G^\dagger(a_j) + \varepsilon_j\right). \tag{1}$$

As the samples are collected through a numerical solution to the RTE on a mesh grid, they will be on a discrete domain $D_j = \{x_1, \ldots, x_n\} \subset D$ with $n$ points, where there is a finite collection of input samples $a_j(x)\big|_{D_j} \in \mathbb{R}^{n \times d_a}$ and output samples $u_j(x)\big|_{D_j} \in \mathbb{R}^{n \times d_u}$. Here, $a_j(x)$ are five discrete input fields representing the scattering coefficient ($\mu_s$), absorption coefficient ($\mu_a$), asymmetry parameter ($g$), geometry, and temperature ($T$), and $u_j(x)$ is the discrete output heat flux ($q''$) field.

The FNO architecture, as shown in Fig. 1, consists of a pointwise lifting layer ($P$), four Fourier layers, and a pointwise projection layer ($Q$). First, the lifting layer takes the input to a higher dimensional representation by $v_0(x) = P(a(x))$, where $P: \mathbb{R}^{d_a} \to \mathbb{R}^{d_v}$ is parameterized by a pointwise $1 \times 1$ convolution and $d_v$ is the network width. Then, the Fourier layer consisting of two paths, a local pointwise path and a nonlocal Fourier integral path, is defined as:

$$v_{t+1}(x) = \sigma\left(W v_t(x) + \mathcal{F}^{-1}\left(R_\phi \cdot \mathcal{F}(v_t)\right)(x)\right) \tag{2}$$

where $t$ is the Fourier layer index, $\sigma: \mathbb{R} \to \mathbb{R}$ is a non-linear activation function, $W: \mathbb{R}^{d_v} \to \mathbb{R}^{d_v}$ is a pointwise (local) linear transform, $\mathcal{F}$ and $\mathcal{F}^{-1}$ denote the forward and inverse Fourier transform, and $R_\phi$ defines a pointwise linear transform in Fourier space on each retained Fourier mode. Finally, the projection layer maps the latent representation from the final Fourier layer ($v_T$) to the output space $u(x) = Q(v_T(x))$. In practice, the higher-frequency Fourier modes are truncated to reduce computational cost and emphasize the dominant low-mid frequency features of the solution. In the architecture implemented in this work, the lowest 16 Fourier modes are retained in each spatial dimension, GELU is the non-linear activation function[29], and the network is trained by minimizing the mean squared error using the Adam optimizer[30] with a learning rate of $1 \times 10^{-3}$ and weight decay of $1 \times 10^{-4}$, and a batch size of 16.

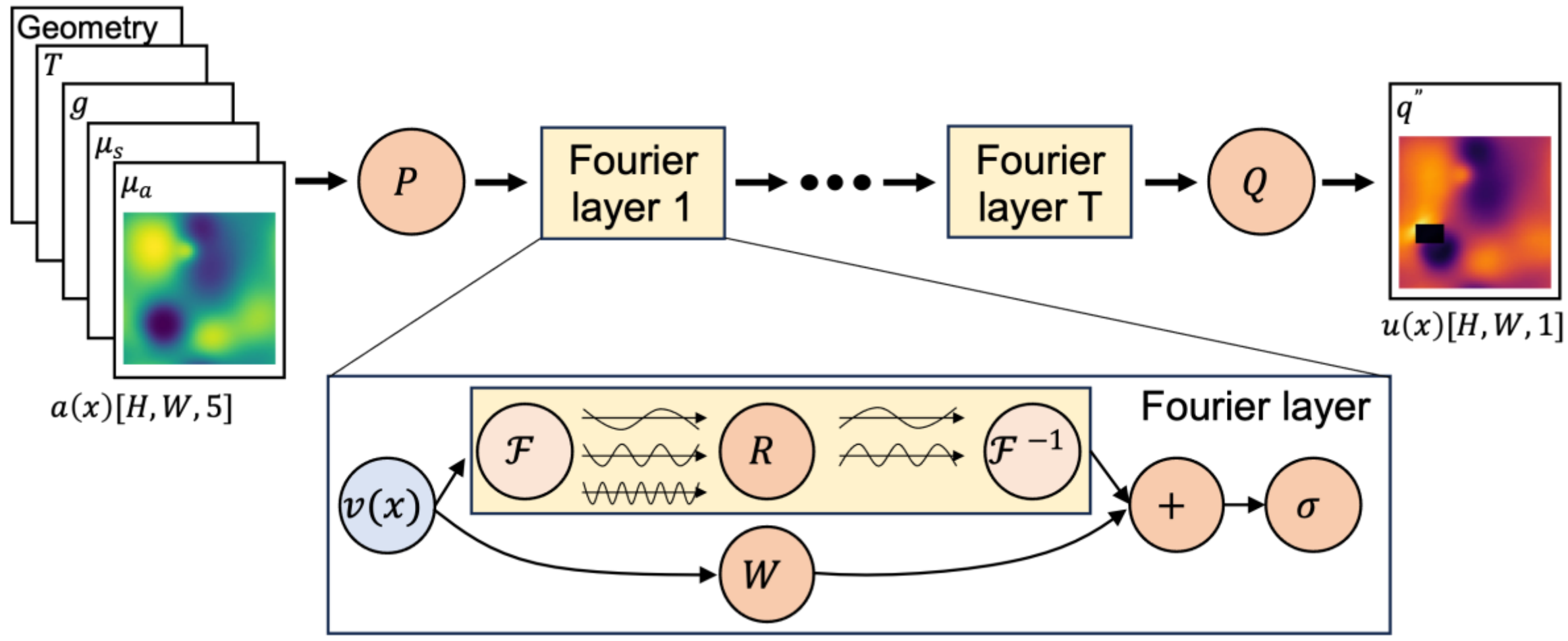

Fig 1: Fourier neural operator architecture with the five discrete inputs $a(x)$, a lifting layer $P$, a series of Fourier layers, a projection layer $Q$, and the discrete output $u(x)$.

*Dataset generation*

A benchmark radiative transfer problem is utilized to train and test the proposed FNO architecture. This problem consists of a square computational domain with periodic boundary conditions applied to all four exterior walls. Each domain includes a solid rectangular element with emitting blackbody surfaces embedded in a participating medium with spatially varying absorption coefficient, scattering coefficient, and asymmetry parameter. For each simulation, the location and dimensions of the rectangle, the rectangle surface temperature, and spatial properties in the participating medium are randomly sampled. The mesh is generated on a Cartesian grid and simulated through a Monte Carlo radiative transport method to generate the absorbed heat flux map within the participating medium.

The spatially varying absorption coefficient, scattering coefficient, and asymmetry parameter are randomly generated as smooth continuous fields. Each property is separately constructed as the superposition of two Gaussian radial basis functions, the first producing a smooth background field, and the second creating localized blobs with different optical properties than the background field. This is calculated as:

$$f(x) = \sum_{k=1}^{N_{bg}} \beta_k exp\left(\frac{-d_k(x)}{2\sigma_k^2}\right) + \sum_{l=1}^{N_{blob}} \beta_l exp\left(\frac{-d_l(x)}{2\sigma_l^2}\right) \quad (3)$$

where $N_{bg}$ is the number of radial basis functions comprising the background field, $N_{blob}$ is the number of random larger blobs, $\beta$ is a random amplitude associated with each basis function, $\sigma$ is the characteristic width, and $d$ is the distance from the basis-function center to the nearest uniformly sampled center. In the dataset generated, $N_{bg} = 80$, $N_{blob} \in [3, 10]$, $\beta_k$ and $\beta_l \sim \mathcal{N}(0, 1)$, $\sigma_k \sim U(0.02, 0.12)$, and $\sigma_l \sim U(0.03, 0.20)$. Finally, the field is normalized by:

$$\theta(x) = \theta_{min} + (\theta_{max} - \theta_{min}) \frac{1}{1 + \exp(-f(x))} \quad (4)$$

where $\theta$ denotes either $\mu_a$, $\mu_s$, or $g$ with $\theta_{min}$ and $\theta_{max}$ being the minimum and maximum values for each property ($0.01 \leq \mu_a \leq 5\ m^{-1}, 0.01 \leq \mu_s \leq 5\ m^{-1}, 0 \leq g \leq 0.95$). A unique field is generated for each spatially varying property. The temperature is a constant value field with $T \sim U(180, 420)\ K$, and the rectangle size is uniformly sampled between 12% and 30% of the domain length ($1\ m$) in each direction. This comprises the five input fields, including the absorption coefficient, scattering coefficient, asymmetry parameter, geometry, and temperature and the output field as the absorbed heat flux, with one example simulation shown in Fig. 2. This process for generating the background field is arbitrary, and selected to evaluate the FNO on a spatially varying participating medium comprised of smooth random features.

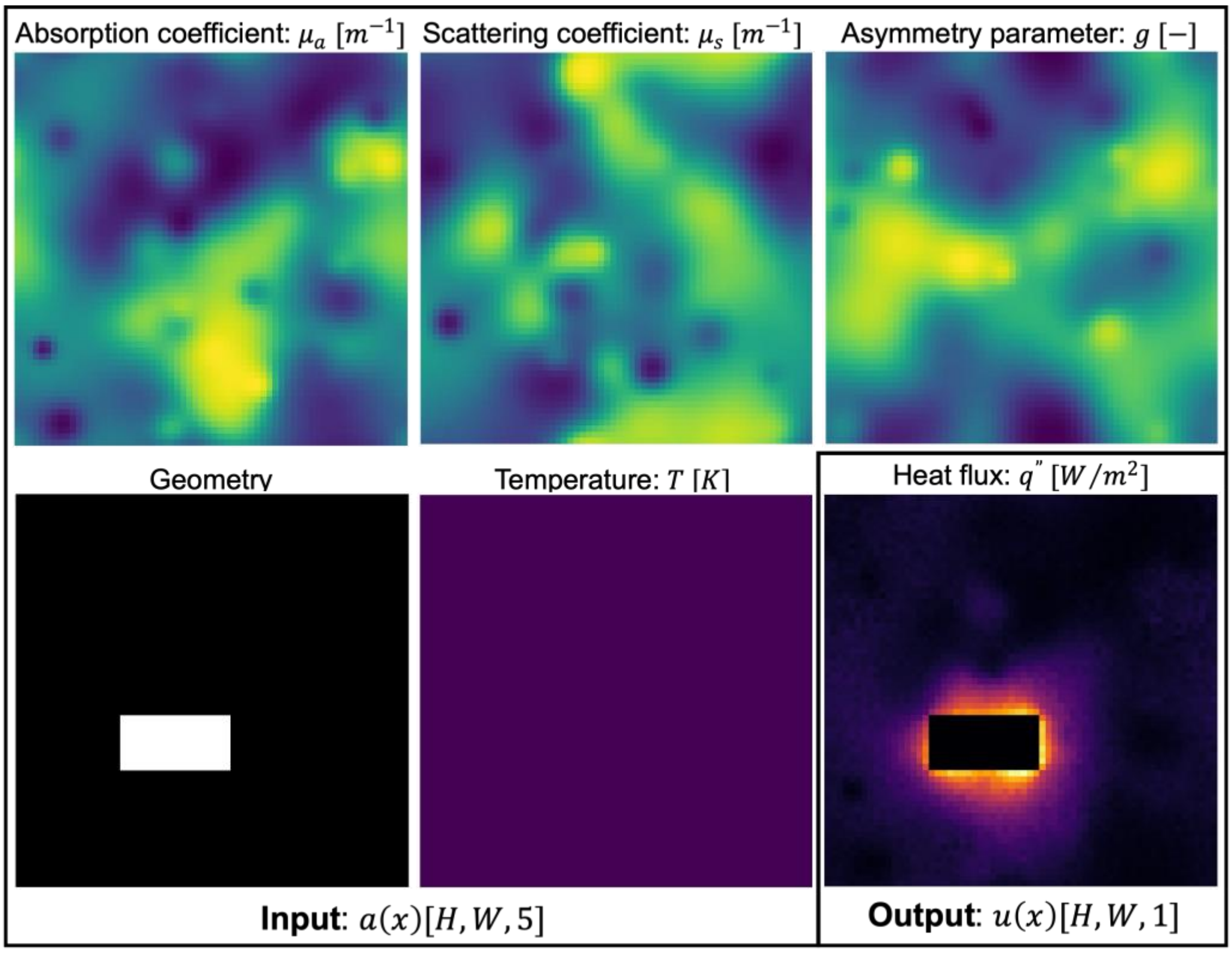

Fig 2: Example individual simulation from the training dataset, including the five input fields and the one output field.

*Monte Carlo method for solving the radiative transfer equation*

Train and test datasets for the FNO are generated through a Monte Carlo method for modeling photon transport in two-dimensional scattering and absorbing media. Photons are launched from uniformly sampled locations and initial propagation angles on the rectangle's surface. Each photon is assigned an initial weight of 1 ($w_0 = 1$) with results normalized afterwards by dividing by the total number of photons sampled ($N$). The photon step size ($s$) is calculated using the Woodcock tracking algorithm[31,32], also referred to as delta tracking or the null-collision algorithm, to aid in handling spatial variation in the optical properties:

$$s = -\ln(\xi)/\mu_{t,max} \tag{5}$$

where $\xi$ is a uniformly distributed random number on the interval (0, 1) and $\mu_{t,max}$ is the maximum extinction coefficient present in the system ($\mu_{t,max} = \max(\mu_s + \mu_a)$). After each step, the probability of a true interaction occurring is defined as the extinction coefficient at the photon's location over the maximum extinction coefficient:

$$p_{real} = \mu_t / \mu_{t,max}. \tag{6}$$

If an interaction does not occur, it is treated as a null collision and the photon continues propagation without updating its weight or direction. If an interaction does occur, the photon weight is reduced due to absorption using the implicit capture algorithm[33,34]:

$$w_{new} = w\left(1 - \frac{\mu_a}{\mu_t}\right). \tag{7}$$

where the absorbed weight is accumulated on a Cartesian grid. Next, the remaining photon weight is scattered, with the updated directional angle sampled from the Henyey-Greenstein phase function distribution using the local asymmetry parameter[35,36]. Although photon transport is modeled in 2D in this study, the scattering angle is sampled from the conventional 3D Henyey-Greenstein distribution. To improve computational efficiency, a Russian roulette algorithm is applied which eliminates photons that have been reduced to small weights while preserving total energy across all photons. Unless otherwise noted, the training dataset is modeled with 500,000 photons per simulation, and the test dataset is modeled with 5,000,000 photons per simulation. This Monte Carlo algorithm is written in Python, with all timed functions compiled to machine code with the Numba just-in-time compiler[37]. Timing is performed after an initial warmup of the algorithm so that compilation time is not included in the reported Monte Carlo time.

*Spectral energy calculation*

To quantify the spatial frequencies present in both the Monte Carlo solutions and the FNO predictions of the heat flux, a discrete Fourier transform is applied to calculate the frequency spectrum present:

$$\hat{u}(k) = \sum_{x} u(x) e^{-2\pi i (k \cdot x)} \tag{8}$$

where in this case, since the dataset is generated in 2-dimensions, $x = (x_1, x_2)$ denotes the spatial coordinates and $k = (k_1, k_2)$ denotes the spatial frequency vector. This is then converted to the radially averaged spectral energy to produce a one-dimensional annular representation of the frequencies present within discrete radial bands:

$$S(k) = \sum_{k \le \|k\| < k + \Delta k} |\hat{u}(k)|^2 \tag{9}$$

where $\Delta k = 1$ is used in this work. Finally, the radially averaged energy spectrum is normalized by $\tilde{S}(k) = S(k) / \sum_k S(k)$.

## Results & Discussion

The FNO performance is first examined on a test dataset of 100 Monte Carlo simulations in Fig. 3 as a function of dataset size, FNO width, and photon count used during training dataset generation. Unless otherwise noted, all other parameters are held constant at an FNO width (W) of 64, training dataset size (D) of 320 simulations, image input and output size of $64 \times 64$, the number of photons sampled in the training dataset at 500,000, and the number of photons sampled in the test dataset at $5 \times 10^6$. Figure 3(a) shows the test loss (mean-squared error) as a function of training dataset size varied from 10 to 1,280 simulations. With dataset size ranging from 10 to 120 simulations a strong power law scaling correlation is observed, consistent with a data-limited regime in which additional samples predictably improve model performance. Beyond a dataset size of approximately 120 simulations, the test loss saturates, indicating a transition to a model size limited regime where further increases in dataset size yield diminishing returns. This type of scaling power law is commonly seen in machine learning models when the varied parameter is the limiting factor[38]. Figure 3(b) varies the FNO width, effectively the number of parameters in the model, from 1 to 64. A brief power law region is seen from widths of 1 to 4, corresponding to a model size limited regime. Increasing the model size beyond approximately 8 yields minimal additional improvement, indicating the model becomes dataset size limited. Joint increases in both dataset size and model width can therefore produce predictable further increases to model accuracy. Figure 3(c) shows the test loss curve across epochs for training datasets with various photon counts. At low photon counts, the test loss initially decreases then increases at later epochs due to overfitting to stochastic Monte Carlo noise. As photon count increases, this behavior collapses and curves approach a similar test loss beyond photon counts of 125,000. This is well below the test dataset photon count of 5 million photons, which may be due to the FNO's ability to filter high-frequency noise provided that label variance is sufficiently small.

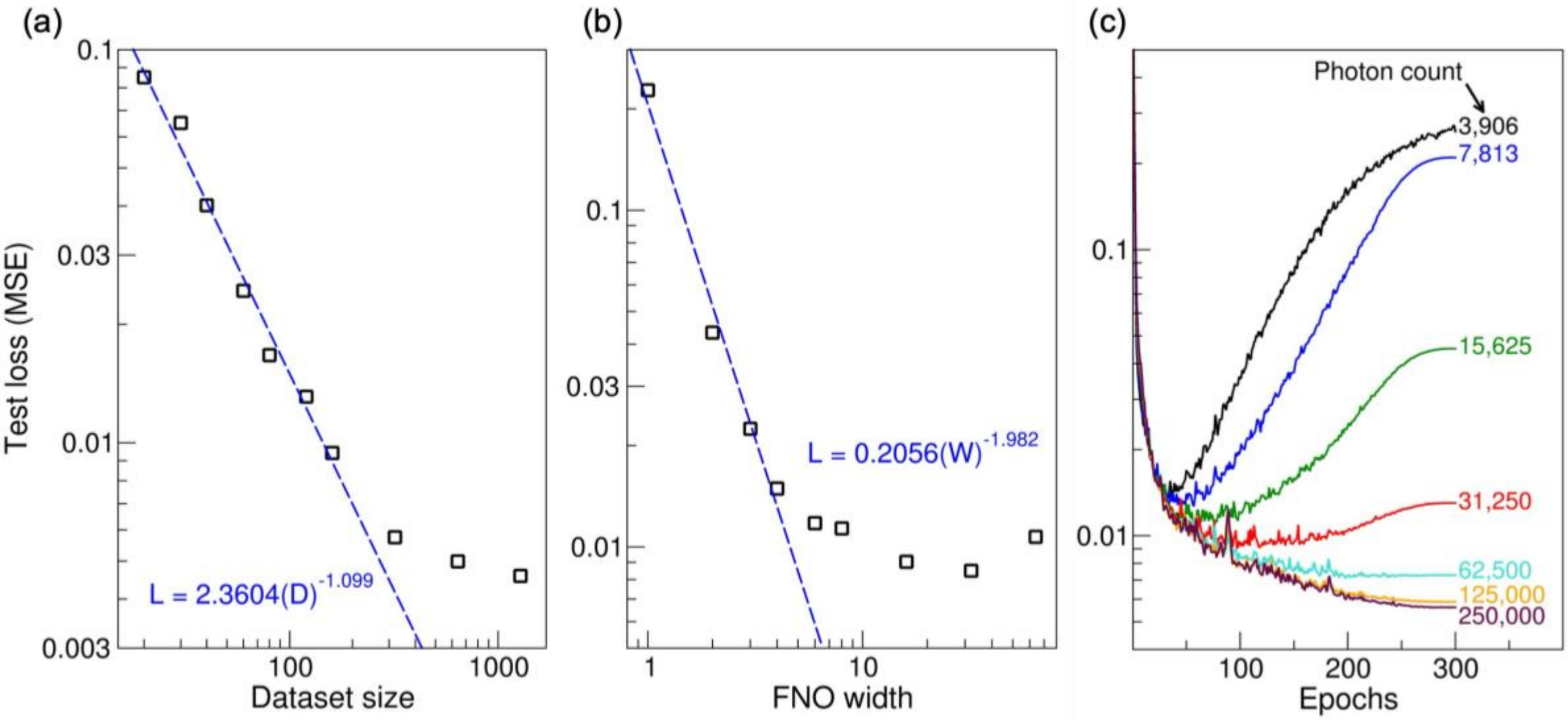

Fig 3: Mean squared error loss on the test dataset as a function of (a) dataset size, (b) FNO width, and (c) epochs at various Monte Carlo photon counts in the training dataset.

The FNO computational cost and prediction accuracy is compared against Monte Carlo simulation in Fig. 4. Figure 4(a) shows the mean absolute error on a test dataset as a function of time per simulation between the FNO at various widths and Monte Carlo with various photon counts. To provide a similar comparison basis, the FNO is implemented in using PyTorch[39], and the Monte Carlo method is implemented in Python with Numba[37] compiling all timed functions to machine code. Both methods are called once prior to timing begins to ensure Just-In-Time compilers are not included in the simulation compute time and are timed in series on a CPU. As expected, the Monte Carlo error scales approximately inversely with the square root of the photon count ($1/\sqrt{N}$). The FNO predictions achieve substantially lower error at a comparable computational cost to Monte Carlo, or alternatively, substantially lower computational cost at a comparable error. Similar to Monte Carlo, a tradeoff between computational power and decreased error is seen by increasing the FNO width (W) from 4 to 16 to 64. The individual FNO predictions at the same width are arranged vertically in Fig. 4(a) because the time required to make a prediction is not dependent on the optical property inputs, whereas Monte Carlo simulations vary in the time required based on the random photon paths sampled and their dependence on the optical property fields. Figure 4(b) summarizes this comparison by plotting the FNO acceleration compared to Monte Carlo simulations at comparable error as a function of FNO width. While the acceleration varies based on FNO width, a peak acceleration of 26× is seen compared to Monte Carlo. These results demonstrate that, after training, the FNO can approximate Monte Carlo-quality radiative transfer solutions at a fraction of the computational cost, making it a promising surrogate for rapid dataset generation and repeated evaluation.

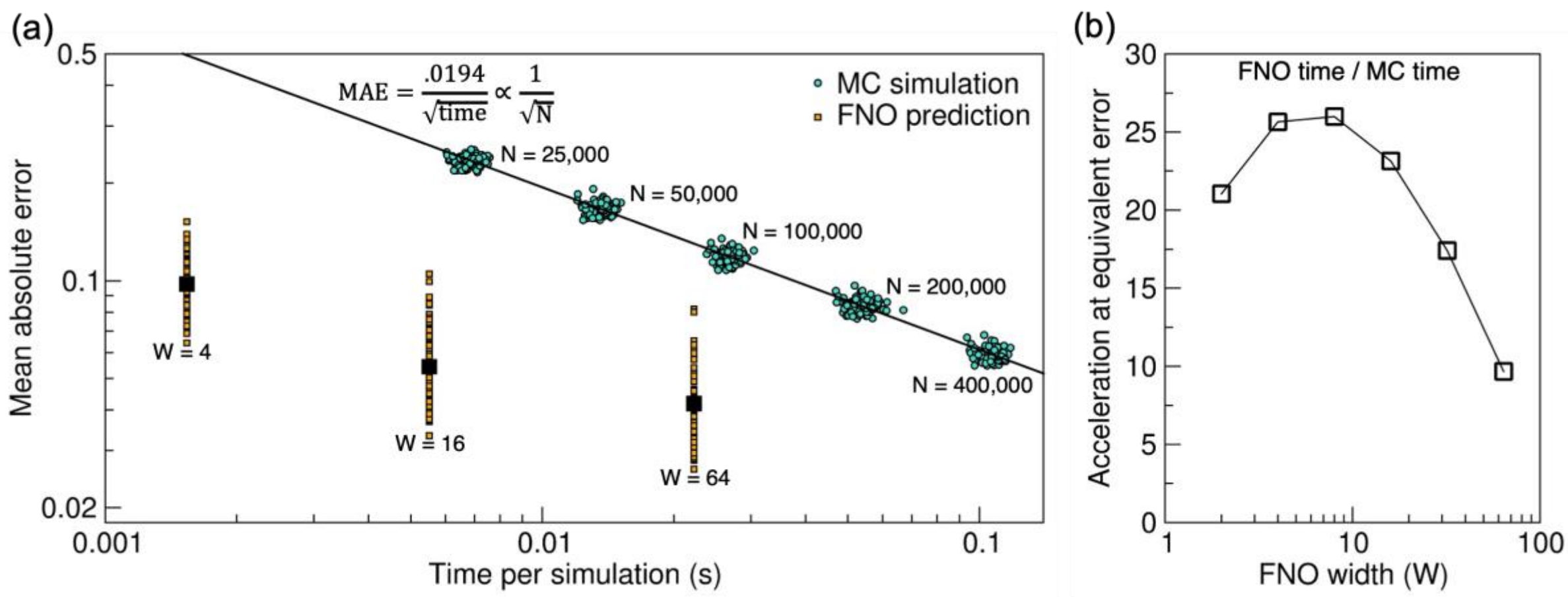

Fig. 4: (a) mean absolute error on the test dataset as a function of time per simulation for Monte Carlo at various photon counts and the FNO at various widths. (b) Acceleration provided by the FNO compared to Monte Carlo at equivalent error as a function of the FNO width.

Figure 5 evaluates the discretization-invariant behavior of the FNO by comparing the test loss as a function of the test dataset grid resolution from 32 × 32 to 512 × 512 with the FNO trained on a single discretization. To evaluate the FNO performance, the number of photons sampled in the test dataset must scale with the number of grid cells to retain an equivalent error due to Monte Carlo noise. For the results seen in Fig. 5(a) only, a unique test dataset is used where the number of photons sampled is 122 photons per grid cell. A portion of the error reported comes from the Monte Carlo noise, however, this error is constant between discretizations due to the increasing photon count allowing for a relative comparison of performance on various discretizations. As shown in Fig. 5(a), the FNO trained on a grid resolution of $32^2$ performs well at $32^2$ but sees significantly increasing prediction error at higher grid resolutions. However, FNO models trained on higher grid resolutions ($64^2$, $128^2$) see a significantly lower increase in the error prediction when tested on higher grid resolutions. This indicates that while the FNO can be utilized on resolutions different from the training resolution, the training dataset resolution must be sufficiently refined to capture the underlying features. Additionally, conservation of energy is an important metric when quantifying the performance of a radiative transfer numerical method. In Fig. 5(b), the percent error in the total predicted energy compared to the true energy emitted from the rectangle surface, defined as $|q_{pred} - q_{emit}|/q_{emit}$, is shown as a function of the grid resolution. A similar trend is seen here, with the FNO trained on a grid resolution of $32^2$ dramatically increasing in error when tested on higher resolutions, and the FNO trained on a grid resolution of $128^2$ achieving limited error across all grid resolutions with a maximum error of 0.63%. In this model, the total predicted energy is not enforced during training or normalized in any post-processing step. These methods could be applied to better enforce energy conservation. To highlight the discretization-invariance of the FNO, an absorption heat flux map from the test dataset is shown in Fig. 6, with an FNO trained on a grid resolution of $64^2$ making predictions on $32^2$, $128^2$, and $512^2$ resolutions, showing excellent agreement with the Monte Carlo ground truth.

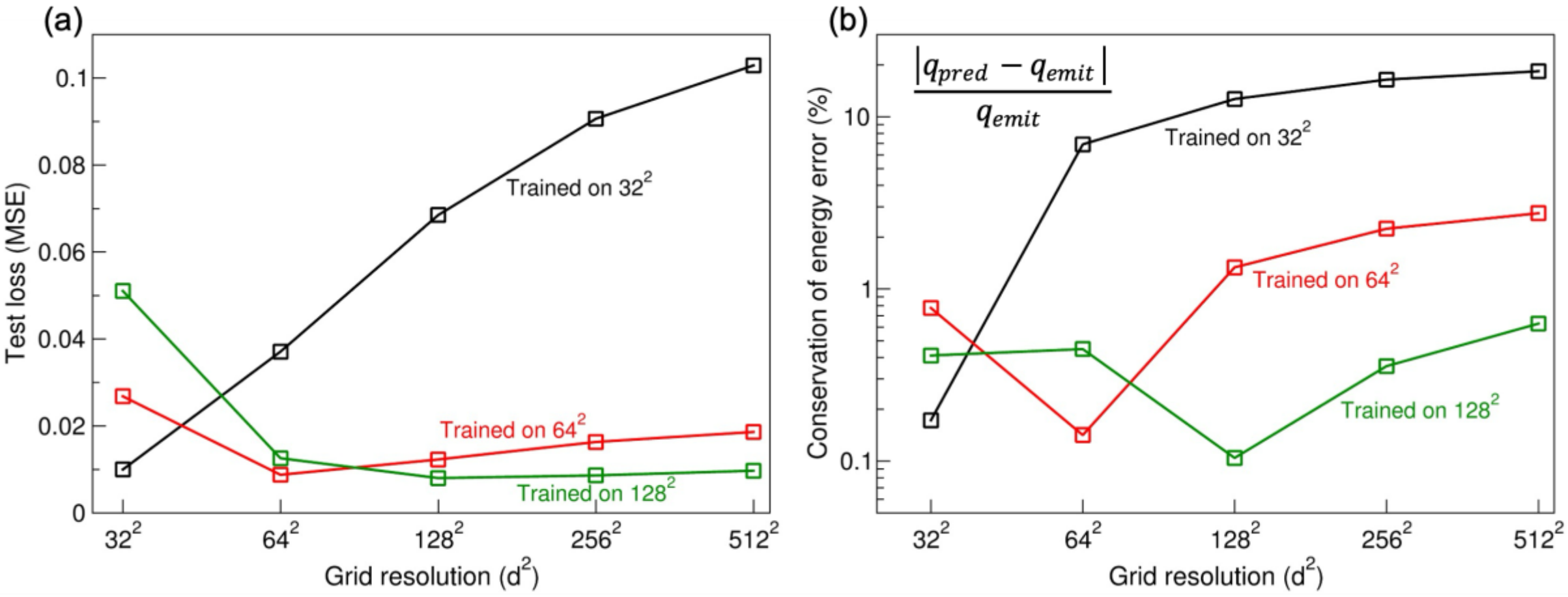

Fig. 5: (a) Mean squared error of the FNO as a function of the test dataset's grid resolution for three FNOs trained only on either $32^2$, $64^2$, or $128^2$ discretization. (b) Percent error in the predicted energy relative to the true energy emitted by the rectangle as a function of the test dataset's grid resolution.

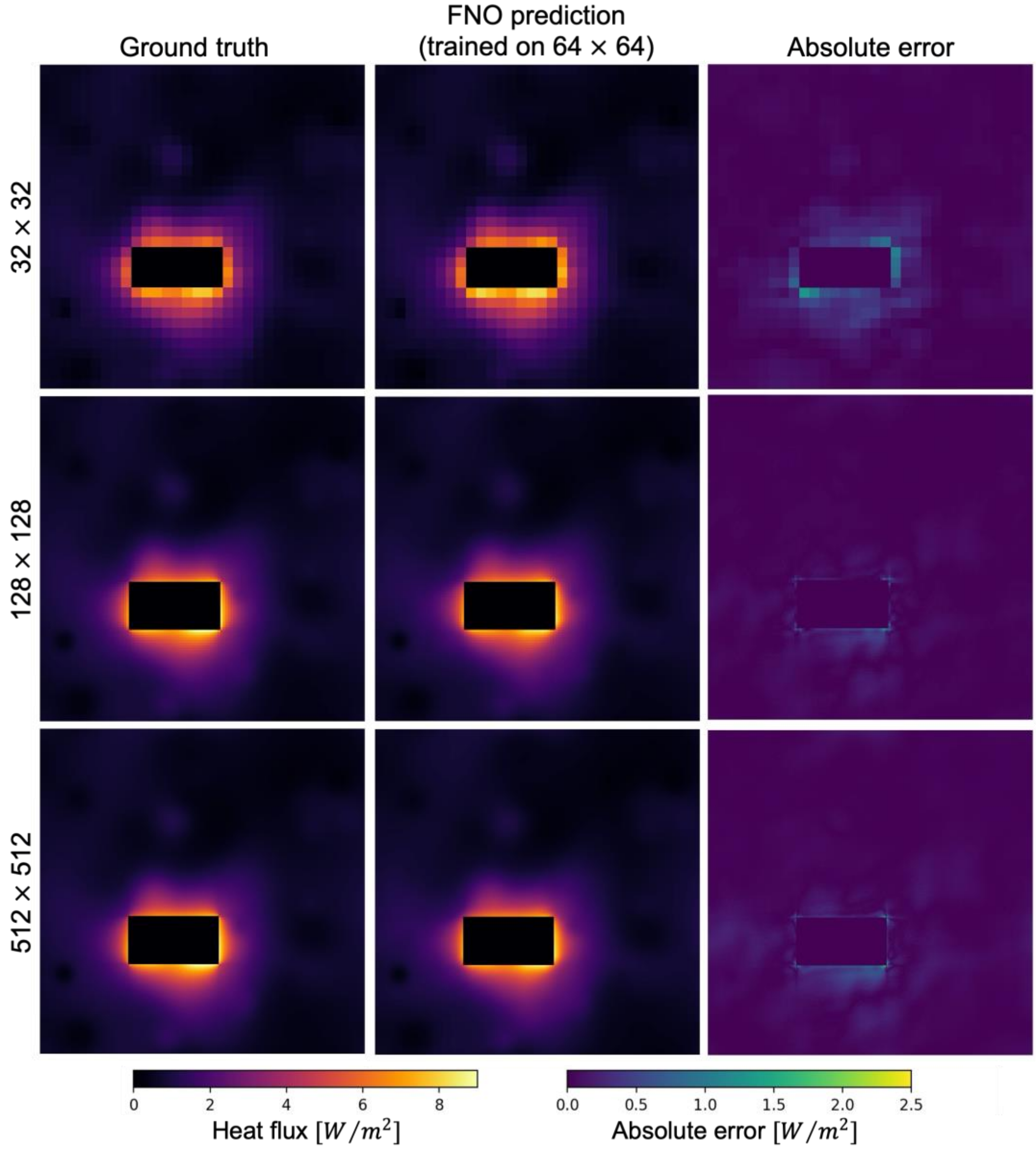

Fig 6: Qualitative single example from the test dataset showing the ground truth at three different discretizations using Monte Carlo with $5 \times 10^6$ photons sampled, the FNO prediction which is solely trained on 64 × 64 discretization, and the absolute error between the FNO prediction and the ground truth.

Figure 7 shows the radially averaged spectral energy to evaluate the spatial frequencies observed in both the Monte Carlo method and the FNO prediction on the test dataset with a 64 × 64 discretization. Here, the FNO is trained on 500,000 Monte Carlo with 500,000 photons. With this discretization, the maximum frequency observed along the length and width is 32, and the maximum frequency observed along the diagonal is approximately 45. This is seen in Fig. 7 with the shaded gray region, the Nyquist region, highlighting the pixel-by-pixel frequencies present in the solution which is where Monte Carlo noise is expected to be seen. In Fig 7(a), the spatial frequencies of Monte Carlo simulations with $5 \times 10^5$, $2.5 \times 10^6$, and $1 \times 10^8$ photons are plotted. In the low-mid frequencies before the Nyquist region, these spatial frequencies are nearly identical, whereas the frequencies present in the Nyquist region decreases with increasing photon count due to the reduced noise associated with higher Monte Carlo photon count. The FNO prediction in low-mid frequencies is nearly identical to the frequencies present in Monte Carlo, however, the FNO predicts significantly lower high-frequency energy than the Monte Carlo solutions. In the benchmark problem evaluated here, the only region with discontinuities where pixel-by-pixel frequencies are expected to appear in the solution are the cells directly adjacent to the solid rectangle. In Fig. 7(b), the same plot is replicated but the cells adjacent to the rectangle in each of the simulations in the test dataset are removed. In the low-mid frequencies the same trend is seen, with all three Monte Carlo photon counts and the FNO providing nearly identical solutions. In the high-frequency Nyquist region, the FNO prediction contains spectral energy levels comparable to that of Monte Carlo with $1 \times 10^8$ photons sampled even though the FNO is only trained on Monte Carlo with $5 \times 10^5$ photons sampled. This shows that while the FNO is not specifically trained to denoise, the FNO architecture demonstrates the ability to learn the true solution operator rather than learning the realization-specific Monte Carlo solutions containing noise. This is not to claim the FNO provides a similarly accurate solution as Monte Carlo with $1 \times 10^8$ photons, but simply that the FNO does not overfit to the noise present in the training dataset. At known discontinuities such as geometry boundaries or rapid changes in volumetric optical properties, the FNO architecture struggles to capture these high-frequency events.

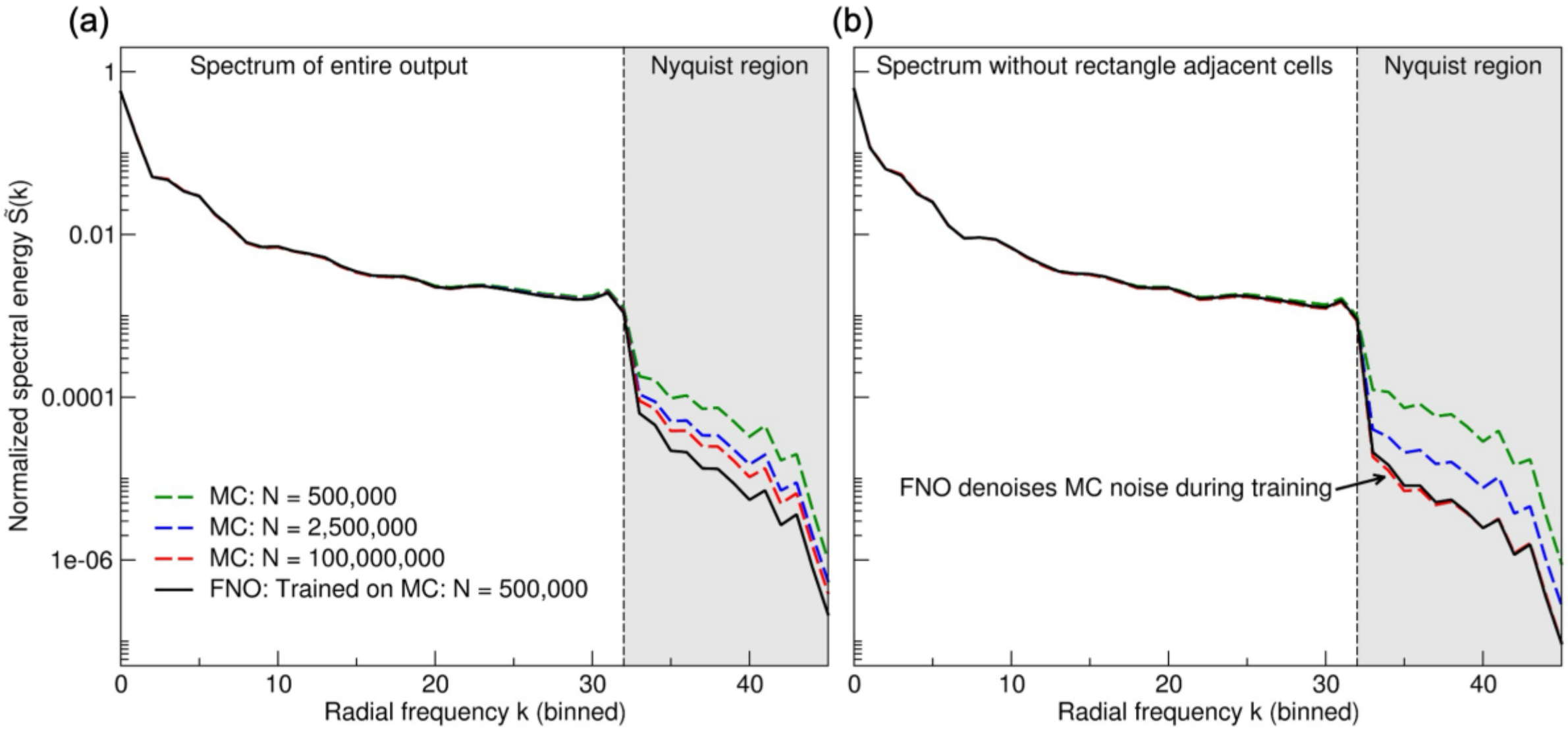

Fig. 7: Normalized spectral energy as a function of the radial frequency for Monte Carlo with three different photon counts and the FNO prediction on (a) the entire output heat flux field and (b) the output heat flux field with the cells directly adjacent to the solid rectangle removed.

Based on the strong performance shown by the FNO architecture here for solving the radiative transfer equation, there are several promising future directions. First, instead of solely inputting and outputting spatial properties and quantities, the network architecture should account for both spatial and angular inputs and outputs, such as radiative source terms and boundary conditions as inputs, and intensity as the output. This will allow for a broad range of boundary conditions (diffuse, specular, coherent source terms) to be applied and various quantities to be evaluated through post-processing. Second, as the radiative transfer equation is a linear Boltzmann equation, radiative transfer from various sources can be superimposed which is not commonly applicable to other partial differential equations with important engineering applications such as fluid dynamics. This means that a single network does not need to learn how to account for all types of source and emission terms. For example, one network could account for emission from surfaces while a separate network could account for volumetric emission. By separating these networks into smaller individual predictors, then superimposing the radiative intensity, the model size, training data, and compute required may be reduced. Finally, the current FNO trained on Monte Carlo simulations does not incorporate physics or provide error estimation. By training a model to predict both the spatial and angular discretized radiative intensity, the discrete ordinates method can be applied both to enforce physics and calculate the prediction residual.

## Conclusion

Numerical solutions to the radiative transport equation are critically important for many applications, however due to the inherently high-dimensional nature of radiative transfer, current solution methods require significant computational cost. In this study, a Fourier neural operator architecture is applied as a discretization-invariant surrogate model for the radiative transfer equation in heterogenous participating media. A dataset is generated on a benchmark 2-dimensional problem with emission from a random rectangular source into a medium with spatially varying properties. The Monte Carlo method is selected to generate absorbed heat flux fields for the dataset due to its unbiased stochastic error. The performance of the FNO is evaluated through scaling the training dataset size, FNO width, and number of photons sampled in Monte Carlo to generate the training dataset. The resulting model provides up to 26× acceleration over Monte Carlo simulation at equivalent error, as well as providing high accuracy on a range of grid discretizations, including discretizations not included in the training dataset. This highlights the FNO's ability to learn the underlying solution operator and not simply an image-to-image mapping. A spectral analysis of the spatial frequencies present shows the FNO preserves the physically meaningful low-mid frequency energy present in the solution, while suppressing the high-frequency energy. This makes the FNO architecture exceptionally efficient at denoising the training dataset which contains high-frequency noise from Monte Carlo, however, does result in error near discontinuities such as geometry boundaries. Overall, these results show the Fourier neural operator architecture provides accurate, computationally efficient, and discretization-invariant predictions for accelerated radiative transport modeling. This approach offers a promising foundation for surrogate models that can substantially reduce the computational cost, especially for repeated evaluations of the radiative transfer equation such as coupled thermal-fluid problems and optimization.

## Declarations

*Competing interest*
The authors declare no competing interests.

*Disclaimer*
The views expressed in this article are those of the author(s) and do not reflect the official policy or position of the U.S. Naval Academy, Department of the Navy, the Department of War, or the U.S. Government.